\documentclass[12pt,preprint]{aastex}
\usepackage{graphicx, amsmath, amssymb}
\usepackage[dvips]{color}
\usepackage{lineno} 
\usepackage{hyperref}

\shortauthors{Troland, et al.}
\shorttitle{Conference Summary: The Cosmic Agitator}

\begin{document}

\title{Conference Summary: \\ The Cosmic Agitator $-$ Magnetic Fields in the Galaxy}

\author{T.~H.\ Troland\altaffilmark{1}, C.\ Heiles\altaffilmark{2},
A.~P.\ Sarma\altaffilmark{3}, G.~J.\ Ferland\altaffilmark{1},
R.~M.\ Crutcher\altaffilmark{4}, C.~L.\ Brogan\altaffilmark{5}}

\altaffiltext{1}{University of Kentucky, Lexington KY}

\altaffiltext{2}{University of California, Berkeley CA}

\altaffiltext{3}{DePaul University, Chicago IL}

\altaffiltext{4}{University of Illinois, Urbana-Champaign IL}

\altaffiltext{5}{National Radio Astronomy Observatory, Charlottesville VA}

\bigskip

\bigskip

\section{INTRODUCTION}
\label{sINTRO}
The very first observations of the galactic magnetic field were made in 1948, 
independently by William Hiltner of Yerkes Observatory and by John Hall of 
Amherst College and the U. S. Naval Observatory.  Both observed polarization 
of starlight, and both published their results in the journal \textit{Science} 
in 1949. In 2008 Mar 26-29, about 80 astronomers met in the scenic city of 
Lexington, Kentucky to celebrate 60 years of observations of the interstellar 
magnetic field, and discuss and debate relevant theoretical studies. The  
conference was formally titled: ``The Cosmic Agitator:
Magnetic Fields in the Galaxy.'' A coincidental reason
for the conference was to celebrate the 60th birthday of Thomas H. Troland
and honor his contributions to the observations of magnetic fields in
the interstellar medium. Troland's birthday provided a serviceable excuse to 
celebrate magnetic fields with a product that goes back much further even than 
Hiltner and Hall --- fine old Kentucky bourbon from Buffalo Trace Distillery. 
Contrary to popular imagination, ``buffalo trace'' is not a waste product from 
the animal.  Instead, it is a wide path established by migrating buffalo that 
were prevalent in Kentucky centuries ago.  The current distillery is built on 
part of what was once such a buffalo thoroughfare. 

In this paper we present a summary of the conference, drawing primarily
from material in the slides prepared for the Conference Summary by one of
us (Carl Heiles). It is ``a necessarily restricted view'' --- restricted
by the 30-minute time slot provided to Heiles for the summary, and the 
number of pages we would have to fill if we tried to summarize the
entire array of wonderful talks and posters. Interested
readers may navigate to the conference web site\footnote{
\href{http://thunder.pa.uky.edu/magnetic/}{http://thunder.pa.uky.edu/magnetic/}}, 
where all the presentations have been posted (except for a few 
presentations whose authors have requested that their talks not be
included on the website).

\section{MAJOR ISSUES: A NECESSARILY RESTRICTED VIEW}
\label{sODR}

The following is an overview of the issues covered at this conference.

\vspace{-0.1in}
\begin{itemize}
		\item \textbf{Observations:}\ There is an overwhelming detail and variety 
	in the observations because of great technical advances over the past 40 years,
	at optical, infrared, millimeter, and centimeter wavelengths. A substantial array
	of measurements	of the perpendicular component of the magnetic field from
	linear polarization observations and the parallel (line-of-sight) component
	of the magnetic field from circular polarization observations is now available.
	
	\item \textbf{Theory:}\ Again, there is an overwhelming amount of theoretical
	work that addresses magnetic fields due to substantial advances in computational
	capability in the past decade.

	\item \textbf{Masers:}\ Four maser transitions (OH, H$_2$O, CH$_3$OH, SiO) allow
	measurements of the magnetic field in star forming regions, supernova shocks 
	and late-type stars in our Galaxy, and even in the extragalactic environment.
	
	\item \textbf{Some specific issues:}\ The morphology of the global Milky Way 
	field remains unclear. Other specific issues addressed relate
	to the containment of overpressured slabs (Galactic and extragalactic), 
	grain alignment physics, and the theory and observation of ambipolar diffusion.
	
	\item \textbf{Instrumentation:}\ Without instrumentation, of course, none of 
	the above would be possible. Several speakers at the conference 
	addressed directly the scope and opportunities from new, upcoming, and
	planned telescopes.
\end{itemize}

\section{LARGE-SCALE FIELDS IN THE MILKY WAY}
\label{sLSF}
There appear to be unresolved issues regarding the large scale magnetic field
of our Milky Way Galaxy. 
Jin Lin Han reported that the large-scale magnetic field in our Galaxy reverses
from arm to interarm. Using rotation measures (RMs) for 223 pulsars, together
with data from the literature, \citet{han2006}\ found evidence for large-scale 
counterclockwise fields (viewed from the north Galactic pole) in the spiral arms 
interior to the Sun. However, in interarm regions, they found that the 
large-scale fields are clockwise, and proposed that the large-scale Galactic 
magnetic field has a bisymmetric structure with reversals on the boundaries 
of the spiral arms. On the other hand, Marijke Haverkorn reported that the
Galactic magnetic field reverses from arm to arm. Using new Faraday RMs
for 148 extragalactic radio sources behind the southern Galactic plane
together with all available extragalactic and pulsars RMs in this region, \citet{bh07}\
generated a simple model for the magnetic field in the fourth quadrant of the 
Milky Way. They find that the magnetic field in the fourth Galactic quadrant 
is directed clockwise in the Sagittarius-Carina spiral arm (as viewed from the 
north Galactic pole), but is oriented counterclockwise in the Scutum-Crux arm
(e.g., see Fig.\ 4 in \citealt{bh07}).  
Are different conclusions being obtained from the same data? Clearly, more 
needs to be done about this important topic; in particular, more data on the
first quadrant is required.

\citet{han1997}\ found that the RMs for extragalactic sources reveal a striking
antisymmetric pattern about the Galactic plane; this is also shown in the
RMs of nearby pulsars at high latitudes. Such a pattern is consistent with
the field configuration of an A0 dynamo. Ann Mao and collaborators have undertaken
a study of the vertical and horizontal components of the magnetic field above 
and below the location of the Sun via a 1.4 GHz survey of about 1000 polarized 
sources ($|b| > 77\arcdeg, L > 4$ mJy). They find a vertical
field of 0.10 $\pm$ 0.02 $\mu$G. Heiles adds the following, which he considers the
most important point: the field direction from the RM fit \textit{agrees} with the 
field orientation from the stellar polarization!

\section{SMALLER SCALE MAGNETIC FIELDS}
\label{sSSF}
Magnetic fields on smaller scales are related to those on larger scales and the
outer turbulence scale. Giles Novak discussed dust polarimetry measurements of 
magnetic field morphology in molecular clouds. He showed that the field 
orientation in four clouds is parallel to the Galactic plane (\citealt{li2006}). 
By comparing SPARO observations of large-scale giant molecular cloud fields with 
simulations of turbulence, he and his collaborators inferred that the magnetic energy 
is comparable to the turbulent energy. Marijke Haverkorn reported that energy 
injection in the spiral arms occurs by multiple mechanisms on size scales ranging
down to pc, not only by single and multiple supernovae at $\sim$100 pc
scales as usually assumed.
The interarm regions do show additional turbulence driving sources up to 
$\sim$100 pc. Energy input in the spiral arms is likely dominated by H II
regions, stellar winds, or protostellar outflows.

\section{THEORY AND SIMULATIONS: NEW DEVELOPMENTS}
\label{sTSD}
Magnetic fields have important dynamical effects from the diffuse to
the dense ISM, as reported by Eve Ostriker. Ostriker discussed the many 
contributions from magnetic effects suggested by theory; for example,
magnetic tension and differential rotation produce turbulence via 
magnetorotational instabilities, etc., and reviewed recent numerical
results related to them. 

Zhi Yun Li discussed magnetic braking and protostellar disk formation.
He noted that the formation of rotationally supported disks is not
guaranteed, and magnetic decoupling together with some other causal
agent is required for disk formation in cores with a realistic 
mass-to-flux ratio ($\lambda$). Simulations show that ambipolar 
diffusion in its simplest form does not enable disk formation for 
realistic levels of cloud magnetization and cosmic ray ionization. 

Ellen Zweibel discussed ambipolar diffusion in a turbulent medium,
and reported that in a turbulent, weakly ionized medium 
magnetic flux is transported relative to the neutrals via
small scale ion-neutral drifts depending on the eddy rate. This 
enhanced diffusion contributes 
to the flat B-field vs.\ density relation in the diffuse ISM 
and it may account for relatively weak fields in molecular clouds.

William Henney discussed the effect of magnetic fields on the evolution
of H II regions. He showed an eerily realistic simulation of an
H II region with magnetic fields (the image is available in Henney's
posted talk on the conference website), which gives an irregular 
H II region boundary which is also permeable --- thus providing 
photon escape paths to form the Warm Ionized Medium (WIM).

Daniel Price discussed the effect of magnetic fields on ISM morphology. 
Even at high $\beta$ where the magnetic energy density is small compared
to thermal and turbulent energy densities, magnetic fields delay and suppress 
star formation. Moreover, it appears that magnetic nature loves a vacuum! 
Strong magnetic fields ($\beta <$ 1) lead to large scale voids, anisotropic 
turbulent motions and column density striations along field lines 
due to streaming motions in the gas. Some of this is demonstrated 
in existing $^{12}$CO observations of Taurus (\citealt{ghb2007}; figure
available in Price's presentation on the Conference website).


\section{PRESSURE OR LIFETIME PROBLEMS}
\label{sTSD}
Tom Troland reported on Zeeman observations of the Orion Veil. The basic 
geometry of the veil from \citet{abel2004} is that of two neutral hydrogen 
sheets (e.g., see their Fig.\ 2) that give rise to components A and B in 
absorption. Troland reported that Component A is dominated by magnetic 
energy whereas component B is in approximate equipartition. In one 
possible scenario, component B lies closer to the trapezium stars, and has
an associated H II region that absorbs the momentum of the stellar UV
radiation. This compresses component B, thereby increasing the magnetic
field and, consequently, the magnetic pressure. The magnetic pressure in 
component B resists the momentum of the absorbed stellar radiation,
putting it in ``hydrostatic'' equilibrium with the stellar radiation field. 

Art Wolfe reported on the detection of an unusually strong magnetic field 
of B$_{los} = 83.9 \pm 8.8 \mu$G in a damped Lyman-$\alpha$ system 
at $z$ = 0.7 toward 3C 286. The neutral gas ($n_e/n < 2 \times 10^{-4}$)
in which this effect has been detected has an extent of over 280 pc, 
and is quiescent, highly magnetized  
($\lambda_n = 2\pi G^{1/2}\Sigma$/$B_{plane}$ = 0.04),
metal-poor and nearly dust-free. Its star 
formation rate per unit area is less than that of our Galaxy.
The detection at $z$ = 0.7 of B$_{los}$ = 83.9 $\mu$G  averaged 
over 200 pc scales is completely unexpected: Dynamo theory predicts 
B-fields  to be weaker, not stronger, in the cosmological past.
Such strong fields are found in  star-forming regions near
centers of galaxies, but the Star Formation Rate in DLA-3C286 is low.
Since magnetostatic equilibrium is not satisfied, a B-field of
$\sim$5 $\mu$G may be enhanced to 100 $\mu$G by a merger-induced shock.
Therefore, an open question is whether a dynamo can build up 
$\sim$5 $\mu$G fields in the 4-5 Gyr age of the disk.

Heiles notes that both the above systems (the Orion Veil and the 
DLA toward 3C 286) appear to present identical problems: a sheet 
of material, highly overpressured, with no apparent means of containment.
Telemachos Mouschovias noted that the magnetically overpressured region 
of the Orion Veil may be confined by the gravitation of the 
Orion Molecular Cloud as a whole.

\section{MASERS}
\label{sM}
Following the presentation by William Watson on masers as a probe of 
magnetic fields, several speakers reported on observations of masers 
in a variety of environments. Crystal Brogan described the observations 
of 1720 MHz OH masers to trace supernova shocks, Anuj Sarma reported 
on observations of H$_2$O masers in star forming regions, Wouter 
Vlemmings reported on methanol masers --- a new addition to the Zeeman 
family, Tim Robishaw reported on Zeeman detections in OH megamasers, 
and Athol Kemball on SiO masers near AGB stars. 

Brogan reported simple Zeeman patterns in 1720 MHz OH masers, yielding
B$_{\theta}$ = 0.2$-$5 mG and weak ($\sim$ 10\%) linear polarization. 
The magnetic pressure is of the order of the shock ram pressure,
and the magnetic field follows the B $\propto (\Delta v)\ n^{0.5}$ (Basu 
relation) \textit{if} the CO line width is used. Moreover, the B-field 
is stronger with higher resolution! Brogan posed the following
(open) questions on this topic: What are the detailed properties 
of the polarization and can we distinguish between theoretical models?
How is the maser flux distributed on small size scales and what are the 
brightness temperatures? Does the B-field really increase with higher 
resolution (which might be indicative of more tangled B-fields on 
larger size scales)? Alas, using VLBA observations with a spatial 
resolution of a few 10s of milliarcseconds, Brogan finds that all of 
the apparent increase in field strength with resolution can be 
explained by spectral blending effects from multiple maser spots 
at slightly different velocities. Brogan's comments on the angle 
made by the magnetic field provided a close interface between 
observations and detailed maser theory described in William 
Watson's presentation.

Methanol masers are the newest addition to the Zeeman family. Wouter
Vlemmings mentioned that they are radiatively pumped in pre-/post-shock regions 
by IR emission from shocks. 
Methanol masers originate in regions where high CH$_3$OH densities are generated 
by evaporation from dust grains, and are often found in similar regions as OH.
Vlemmings reported significant detections in 17 of 24 star forming regions with
line-of-sight field strengths of 12 mG. Field reversals are also detected
in two sources. Heiles adds, however, that what is commonly referred to as "reversals"
are more likely to be small changes in angle to the plane-of-the-sky.
The line-of-sight field directions from Vlemmings' work are consistent with 
OH maser observations, and therefore indicative of the Galactic magnetic field
direction.

Tim Robishaw reported on the detection of magnetic fields in
5 ULIRGs, via observation of the Zeeman effect in OH megamasers. These
are the first extragalactic Zeeman splitting detections in emission lines;
the only previous extragalactic Zeeman detection was in H I absorption lines
in the high-velocity cloud around Perseus A (\citealt{kazes91}; 
\citealt{sarma2005}). The detected B-fields are similar to Galactic
sites of OH masers, which is not surprising since conditions in regions 
of massive star formation are similar to those in the Milky Way. Moreover, the
detected B-fields are consistent with magnetic fields in ULIRGs
inferred from minimum energy and equipartition considerations.

\section{TURBULENCE, AMBIPOLAR DIFFUSION AND OTHER ISSUES}
\label{sTAD}
Alex Lazarian discussed the current state of grain alignment theory 
and its implications. Lazarian remarked that polarization from aligned 
dust allows us to trace magnetic fields if we understand grain alignment.
For example, some apparent reversals in field direction may be due to 
radiative torques (RATs).

Martin Houde seems to be directly detecting ambipolar diffusion, via a
difference in line widths between coexistent ionic and neutral 
species. Ambipolar diffusion and turbulence were discussed by
a number of theorists, including Mordecai-Mark Mac Low, Telemachos
Mouschovias, Shantanu Basu, Fabian Heitsch, and Konstantinos Tassis. For
the ``turbulence-community'' Mac Low stated that the bottom line is 
that magnetic fields are important, but they don't dominate 
molecular core formation. Instead, star formation is regulated by
turbulent flows modulated by magnetic fields. Shantanu Basu discussed the 
issue of core formation due to magnetic fields, ambipolar diffusion 
and turbulence.

Many new and impressive results on dense and star-forming regions were 
reported by Roger Hildebrand, Brenda Matthews, Giles Novak, Dick Crutcher, 
and Ramprasad Rao. Based on (ongoing) statistical analysis of the full 
set of Zeeman results presented in \citet{rmc99}, Crutcher concluded that 
the increase in the mass-to-flux ratio from the envelope to the core 
required by ambipolar diffusion is not seen. However, Telemachos 
Mouschovias questioned whether this result is truly inconsistent 
with the ambipolar diffusion-driven models of star formation.  He 
suggested that a more careful comparison of theory and observations 
is warranted.

A number of graduate students gave impressive talks and poster presentations,
signalling that the next generation is ready and willing to continue on the
path blazed by the seasoned pioneers and expand the frontiers of knowledge. 
Sui Ann Mao's result on the magnetic field above and below the Galactic 
plane, and Tim Robishaw's detection of the Zeeman effect in ULIRGs
have already been discussed above. Talayeh Hezareh reported on a method to 
simultaneously calculate the Cosmic Ray Ionization Rate and Fractional 
Ionization in DR21(OH).

\section{INSTRUMENTATION}
\label{sINSTR}

The future looks bright for instrumentation, if funds allow. 
Optical and IR polarization can now be done with CCDs, instead
of single-pixel polarimeters! Antonio Mario Magalhaes reported on
a proposed 2-3 meter robotic telescope for starlight polarization,
and he has it all designed! Dan Clemens discussed IR polarization, 
and informed us that the Galactic Plane Infrared Polarization Survey 
(GPIPS) is up and running, and will be about 50\% complete after the 
current season. When completed, it is expected to yield about 
400,000 new H-band stellar polarizations across 76 square degrees 
of the inner Galactic Plane. Darren Dowell made the case for an 
instrument performing far-IR polarimetry from space, pointed out 
that no planned mission seems
to have considered this need, and stressed the need to start 
campaigning for a polarimeter in space. Brenda Matthews examined the
prospects for dust linear polarization measurements with SCUBA-2; Heiles
remarked that SCUBA-2 is fantastic!
Following his report on polarimetric studies of interstellar turbulence,
Roger Hildebrand reiterated that SOFIA needs a polarimeter.

Radio and mm wavelengths also offer great promise.
Rick Perley predicted that EVLA's polarization capability will 
ovewhelm us. Dick Plambeck revealed the mouth-watering prospects
of CARMA's upcoming polarization capability. Ramprasad Rao showed
how the SMA is already a fantastic polarimeter! Al Wooten described
how ALMA will be the ultimate mm-wave polarimeter when it arrives.

\section{SUMMARY}
\label{sSUM}
Just as the buffalo grazing through Kentucky left behind a site
that would fulfill the Bacchanalian aspirations of generations of
humanity, so it is our hope that the hordes of astronomers trekking
through the Bluegrass state will leave behind not only a memory of days 
spent in great enjoyment and wonder at cosmic mysteries, but a 
firm foundation on the relevance of the magnetic field to the Universe
upon which to build in future. This conference showed us how far
we have come from the days of Hiltner \& Hall, and gave us an inkling
of how far we have to go. The future appears to be full of promise, and 
on that note of hope we shall end our summary of the proceedings, thereby
allowing thirsty astronomers to return to oversized bottles of
Kentucky bourbon which were (hopefully) not confiscated at airport
check points!

\acknowledgments
We thank all the participants who traveled to the Bluegrass state 
from near and far for a very stimulating and wonderful
conference. Thanks are due to support staff at the University
of Kentucky (UK): Eva Ellis (Dept.\ of Physics \& Astronomy)
and Richard Mullins for responding swiftly and efficiently to all
requests for assistance, and to Clay Gaunce of UK Tech Support and other 
members of the Tech Support team for assistance beyond the call of duty.
We would like to acknowledge the support provided by the UK
College of Arts \& Sciences Graduate School, UK Department 
of Physics and Astronomy and the Center for Computational Sciences, UK. We 
have used extensively the NASA Astrophysics Data System (ADS) astronomy 
abstract service, and the astro-ph web server.



\end{document}